\documentclass[prd,twocolumn,tightenlines,superscriptaddress,nofootinbib,
showpacs]{revtex4}

\usepackage{amssymb,latexsym}
\usepackage{amsmath,amsbsy,bbm}
\usepackage{epsfig,bm}
\usepackage{graphicx,comment}
\usepackage{color}
\usepackage{bbm}
\newcommand{\R}{{\mathbb{R}}}

\newcommand{\p}{\partial}
\newcommand{\ontopof}[2]{\genfrac{}{}{0pt}{}{#1}{#2}}

\begin{document} 

\title{Systematic Effective Field Theory Investigation of Spiral Phases in
Hole-Doped Antiferromagnets on the Honeycomb Lattice}

\author{F.-J.~Jiang}
\email[]{fjjiang@itp.unibe.ch}
\affiliation{Center for Research and Education in Fundamental Physics, 
Institute for Theoretical Physics, Bern University, 
Sidlerstrasse 5, CH-3012 Bern, Switzerland}
 
\author{F.~K\"ampfer}
\email[]{fkampfer@mit.edu}
\affiliation{Condensed Matter Theory Group, Department of Physics, 
Massachusetts Institute of Technology (MIT), 77 Massachusetts Avenue,
Cambridge, MA 02139, U.S.A.}
\author{C.~P.~Hofmann}
\email[]{christoph@ucol.mx}
\affiliation{Facultad de Ciencias, Universidad de Colima,
Bernal D\'iaz del Castillo 340, Colima C.P.\ 28045, Mexico}
\author{U.-J.~Wiese}
\email[]{wiese@itp.unibe.ch}
\affiliation{Center for Research and Education in Fundamental Physics, 
Institute for Theoretical Physics, Bern University, 
Sidlerstrasse 5, CH-3012 Bern, Switzerland}
\affiliation{Institute for Theoretical Physics, ETH Z\"urich, 
Schafmattstrasse 32, CH-8093 Z\"urich, Switzerland}

\vspace{-1cm}

\begin{abstract}
Motivated by possible applications to the antiferromagnetic precursor of the
high-temperature superconductor Na$_x$CoO$_2\cdot$yH$_2$O, we use a systematic 
low-energy effective field theory for magnons and holes to study different 
phases of doped antiferromagnets on the honeycomb lattice. The effective action
contains a leading single-derivative term, similar to the Shraiman-Siggia term 
in the square lattice case, which gives rise to spirals in the staggered 
magnetization. Depending on the values of the low-energy parameters, either a 
homogeneous phase with four or a spiral phase with two filled hole pockets is 
energetically favored. Unlike in the square lattice case, at leading order the 
effective action has an accidental continuous spatial rotation symmetry. 
Consequently, the spiral may point in any direction and is not necessarily 
aligned with a lattice direction. 
\end{abstract}
\pacs{74.20.Mn, 75.30.Ds, 75.50.Ee, 12.39.Fe}

\maketitle

\section{Introduction}

Since the discovery of high-temperature superconductivity in the cuprates 
\cite{Bed86}, identifying the dynamical mechanism behind it remains one of the 
biggest challenges in condensed matter physics. It has been suggested that the 
physics of high-temperature superconductivity can be described by $t$-$J$-type 
models. Using a variety of techniques, numerous interesting properties of doped
antiferromagnets have been investigated in great detail both numerically and 
analytically 
\cite{Hir85,And87,Gro87,Tru88,Shr88,Sch88,Cha89,Neu89,Kan89,Wen89,Fis89,Sac89,Sha90,And90,Sin90,Tru90,Kan90,Els90,Dag90,Cha90,Has91,Aue91,Sar91,Ede91,Arr91,Iga92,Fre92,Has93,Psa93,Mor93,Sus94,Chu94,Chu95,Zho95,Chu98,Kar98,Man00,Bru00}. 
For instance, as was first pointed out by Shraiman and Siggia \cite{Shr88}, a 
spiral phase with a helical structure in the staggered magnetization is a 
candidate ground state of doped antiferromagnets even at arbitrarily small 
doping
\cite{Kan89,Kan90,Cha90,Aue91,Sar91,Ede91,Arr91,Iga92,Fre92,Psa93,Mor93,Chu95,Chu98,Zho95,Kar98,Man00,Sus04,Kot04}. 
Unfortunately, due to the strong electron correlations in these systems, most 
analytic results suffer from uncontrolled approximations. Similarly, numerical 
simulations suffer from a severe sign problem away from half-filling. 
Consequently, although numerous investigations have been devoted to 
understanding the spiral phases in doped antiferromagnets, some controversial 
results have been obtained.

In analogy to chiral perturbation theory for the pions in QCD 
\cite{Wei79,Gas85}, a systematic low-energy effective field theory for the 
magnons in an antiferromagnet was developed in 
\cite{Cha89,Neu89,Fis89,Has90,Has91,Has93,Chu94}. Motivated by the success of 
baryon chiral perturbation theory for pions and nucleons
\cite{Geo84,Gas88,Jen91,Ber92,Bec99}, respecting the symmetry constraints of
the underlying $t$-$J$ model and taking into account the location of the 
hole or electron pockets in momentum space, low-energy effective field theories
for magnons and holes or electrons have been constructed for lightly doped 
antiferromagnets on the square lattice in \cite{Kae05,Bru06,Bru07a}.
The effective theories are universally applicable and yield results that are
exact, order by order in a systematic low-energy expansion. Material-specific
properties enter the effective Lagrangian in the form of a priori undetermined
low-energy parameters, like the spin stiffness $\rho_s$ or the spinwave
velocity $c$. The effective theories for hole- and electron-doped systems were 
used to investigate the one-magnon exchange potentials and the resulting 
two-hole or two-electron bound states as well as possible spiral phases 
\cite{Bru06,Bru06a,Bru07,Bru07a}. In the hole-doped case, the leading order
magnon-hole coupling is described by the Shraiman-Siggia term that contains 
just a single spatial derivative. For sufficiently small $\rho_s$, even at
arbitrarily small hole density, this term stabilizes a zero degree spiral phase
in which the spiral is oriented along a lattice axis. In the electron-doped 
case, on the other hand, the Shraiman-Siggia term is forbidden by the 
symmetries, and, consequently, spiral phases are not energetically favorable.

In addition to the cuprates, another superconducting material, 
Na$_x$CoO$_2\cdot$yH$_2$O \cite{Tak03}, has attracted a lot of attention 
\cite{Bas03,Tru04,Che04,Mil04,Zhe04,Wat05}.
The underlying triangular lattice of this geometrically frustrated 
material leads to a severe sign problem and thus prevents us from studying it
from first principles using Monte Carlo calculations. On the other hand, the 
honeycomb lattice structure of the dehydrated variant of 
Na$_x$CoO$_2 \cdot$yH$_2$O at $x=1/3$ has motivated several investigations of 
the antiferromagnetism as well as the single-hole dispersion relation on the 
non-frustrated honeycomb lattice \cite{Lus04,Jia08}. In particular, the 
low-energy parameters of the effective theory for the $t$-$J$ model, namely the
staggered magnetization $\widetilde{{\cal M}}_s$ \cite{Cas05}, the 
spin-stiffness $\rho_s$, 
the spinwave velocity $c$, and the kinetic mass of a hole $M'$ have been 
determined with high precision using an efficient cluster algorithm 
\cite{Jia08}. 

Motivated by possible applications to Na$_x$CoO$_2 \cdot$yH$_2$O, using the 
same methods as for the square lattice
\cite{Kae05,Bru07a}, we have constructed a systematic 
effective field theory for the $t$-$J$ model on the honeycomb lattice. The 
details of this construction will be presented in a forthcoming publication 
\cite{Kae08}. In this work, we apply the resulting effective Lagrangian to 
investigate possible spiral phases of lightly hole-doped antiferromagnets on 
the honeycomb lattice. In contrast to the square lattice case, the leading 
terms of the effective Lagrangian have an accidental continuous rotation 
symmetry. This implies that possible spirals are not necessarily aligned with a
lattice direction. Assuming that the 4-fermion couplings between holes can be 
treated perturbatively, the effective theory predicts that, depending on the 
values of the low-energy parameters, either a homogeneous phase with four or a 
spiral phase with two occupied hole pockets is energetically favored.

The rest of this paper is organized as follows. In section \ref{1} we review 
the effective theory for magnons and holes in an antiferromagnet on the 
honeycomb lattice. In particular, we list the transformation properties of 
magnon and hole fields under the symmetries of the underlying microscopic 
$t$-$J$ model, and we discuss the accidental spatial rotation invariance of the
leading terms in the effective Lagrangian. In section \ref{2} we consider the 
homogeneous and possible spiral phases restricting ourselves to configurations 
that induce a homogeneous background field for the doped holes. In section 
\ref{3}, we include the 4-fermion couplings using perturbation theory and 
investigate the stability ranges of the various phases. Finally, section 
\ref{4} contains our conclusions.
 
\section{Systematic Low-Energy Effective Field Theory for Magnons and Holes}
\label{1}
In this section we briefly review the effective theory for magnons and holes in
an antiferromagnet on the honeycomb lattice. In particular, we list the 
symmetry transformation rules for magnon and hole fields under the various 
symmetries of the underlying $t$-$J$ model which is essential for constructing 
the effective Lagrangian. The staggered magnetization of an antiferromagnet is 
described by a unit-vector field
\begin{equation}
\vec e(x) =
(\sin\theta(x) \cos\varphi(x),\sin\theta(x) \sin\varphi(x),\cos\theta(x)),
\end{equation}
in the coset space $SU(2)_s/U(1)_s = S^2$, with $x = (x_1,x_2,t)$ 
denoting a point in $(2+1)$-dimensional space-time. A key ingredient for 
constructing the effective field theory is the nonlinear realization of the 
global $SU(2)_s$ spin symmetry which is spontaneously broken down to its 
$U(1)_s$ subgroup \cite{Kae05}. This construction leads to an Abelian ``gauge''
field $v^3_{\mu}(x)$ and to two vector fields $v^{\pm}_{\mu}(x)$ which are 
``charged'' under $U(1)_s$ spin transformations. The coupling of magnons to 
holes is realized through a matrix-valued anti-Hermitean field
\begin{equation}
v_\mu(x) = i v_\mu^a(x) \sigma_a, \qquad 
v_\mu^\pm(x) = v_\mu^1(x) \mp i v_\mu^2(x),
\end{equation}
which decomposes into an Abelian ``gauge'' field $v_\mu^3(x)$ and two vector 
fields $v_\mu^\pm(x)$ ``charged'' under the unbroken subgroup $U(1)_s$. 
Here $\vec \sigma$ are the Pauli matrices. These 
fields have a well-defined transformation behavior under the symmetries which 
the effective theory inherits from the underlying microscopic $t$-$J$ model
\begin{alignat}{2}
\label{symm_magnonfields}
SU(2)_s:&\quad &v_\mu(x)' &= h(x) (v_\mu(x) + \p_\mu) h(x)^\dagger,
\nonumber \\
D_i:&\quad &^{D_i}v_\mu(x) &= v_\mu(x),
\nonumber\\
O:&\quad &^O v_1(x) &=
\tau(Ox)\big(\tfrac{1}{2}v_1(Ox)+\tfrac{\sqrt{3}}{2}v_2(Ox)\nonumber\\
&\quad&\, & +\tfrac{1}{2}\p_1+\tfrac{\sqrt{3}}{2}\p_2\big)\tau(Ox)^\dagger,
\nonumber\\
&\quad &^O v_2(x) &=
\tau(Ox)\big(-\tfrac{\sqrt{3}}{2}v_1(Ox)+\tfrac{1}{2}v_2(Ox)\nonumber\\
&\quad&\, &-\tfrac{\sqrt{3}}{2}\p_1+\tfrac{1}{2}\p_2\big)\tau(Ox)^\dagger,
\nonumber\\
&\quad &^O v_t(x) &=
\tau(Ox)(v_t(Ox)+\p_t)\tau(Ox)^\dagger,\nonumber\\
R:&\quad &^R v_1(x) &= v_1(Rx), \quad
^R v_2(x) = -v_2(Rx), \nonumber \\
&\quad &^R v_t(x) &= v_t(Rx),\nonumber \\
T:&\quad &^Tv_i(x) &= \tau(Tx)(v_i(Tx)+\p_i)\tau(Tx)^\dagger, \nonumber\\
&\quad &^Tv_t(x) &= -\tau(Tx)(v_t(Tx)+\p_t)\tau(Tx)^\dagger,
\end{alignat}
where $D_i$, with $i \in \{1,2\}$, are the displacements along primitive 
translation vectors which are chosen to be
$a_1 = (\frac{3}{2}a,\frac{\sqrt{3}}{2}a)$ and 
$a_2 = (0,\sqrt{3}a)$, respectively. Here $a$ is the lattice spacing. Further, 
$O$, $R$, and $T$ in eq.(\ref{symm_magnonfields}) represent a 60 degrees 
spatial rotation around the center of a hexagon, a spatial reflection, and time
reversal, which are given by
\begin{eqnarray}
Ox&=&O(x_1,x_2,t) = 
(\tfrac{1}{2}x_1 - \tfrac{\sqrt{3}}{2}x_2, \tfrac{\sqrt{3}}{2}
x_1 + \tfrac{1}{2}x_2, t), \nonumber \\
Rx&=&R(x_1,x_2,t) = (x_1,-x_2,t), \nonumber \\
Tx&=&T(x_1,x_2,t) = (x_1,x_2,-t),
\end{eqnarray}
respectively. In expressing these symmetry transformation properties, 
we have introduced the matrix $\tau(x)$ which takes the form
\begin{equation}
\tau(x)=
\left(
\begin{array}{cc}
0 & -\exp(-i \varphi(x))\\
\exp(i\varphi(x)) & 0
\end{array}
\right).
\end{equation}
Finally, the Abelian ``gauge'' transformation
\begin{equation}
h(x) = \exp(i\alpha(x)\sigma_3)
\end{equation}
belongs to the unbroken $U(1)_s$ subgroup of $SU(2)_s$ and acts on the 
composite vector fields as
\begin{eqnarray}
v_\mu^3(x)' &=& v_\mu^3(x) - \p_\mu \alpha(x), \nonumber \\ 
v_\mu^\pm(x)' &=& v_\mu^\pm(x) \exp(\pm 2 i \alpha(x)).
\end{eqnarray}

Analytic calculations as well as Monte Carlo simulations in $t$-$J$-like models
on the honeycomb lattice have revealed that at small doping holes occur in 
pockets centered at lattice momenta $k^\alpha = - k^\beta = 
(0,\frac{4\pi}{3\sqrt{3}a})$, and their copies in the periodic Brillouin zone 
\cite{Lus04,Jia08}. The honeycomb lattice, illustrated in figure 1, is a 
bipartite non-Bravais lattice which consists of two triangular Bravais 
sublattices.
\begin{figure}[t]
\begin{center}
\epsfig{file=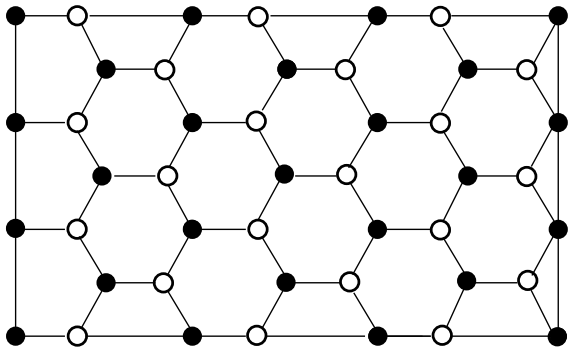,width=7.5cm}
\end{center}
\caption{\it Bipartite non-Bravais honeycomb lattice consisting of two 
triangular Bravais sublattices.}
\end{figure}
The corresponding Brillouin zone and the corresponding hole pockets are shown 
in figure 2.
\begin{figure}[t]
\begin{center}
\epsfig{file=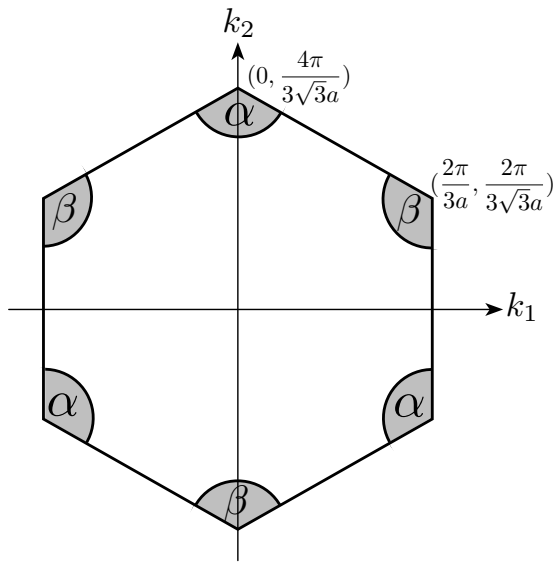,width=7.5cm}
\end{center}
\caption{\it Brillouin zone of the honeycomb lattice with corresponding hole
pockets.}
\end{figure}
The single-hole dispersion relation for the $t$-$J$ model on the honeycomb 
lattice is illustrated in figure 3.
\begin{figure}[t]
\begin{center}
\epsfig{file=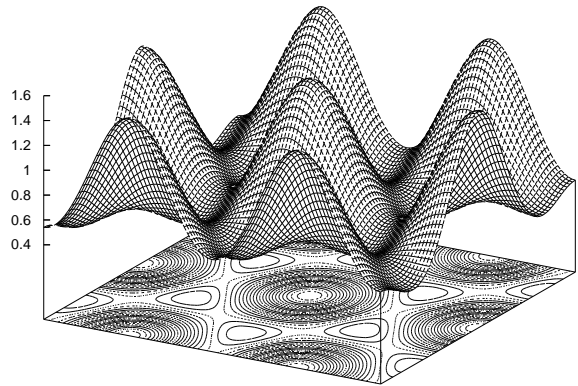,angle=-90,width=7.5cm}
\end{center}
\caption{\it Energy-momentum dispersion relation $E_h(k)/t$ for a single hole 
in the $t$-$J$ model on the honeycomb lattice for $J/t = 2$.}
\end{figure}

The effective field theory is defined in the 
space-time continuum and the holes are described by Grassmann-valued fields 
$\psi^f_s(x)$ carrying a ``flavor'' index $f = \alpha, \beta$ that 
characterizes the corresponding hole pocket. The index $s = \pm$ denotes spin 
parallel ($+$) or antiparallel ($-$) to the local staggered magnetization. 
As will be shown in \cite{Kae08}, under the various symmetry operations the 
hole fields transform as
\begin{alignat}{2}
SU(2)_s:&\quad &\psi^f_\pm(x)' &= \exp(\pm i \alpha(x)) \psi^f_\pm(x),
\nonumber \\
U(1)_Q:&\quad &^Q\psi^f_\pm(x) &= \exp(i \omega) \psi^f_\pm(x),
\nonumber \\
D_i:&\quad &^{D_i}\psi^f_\pm(x) &=
\exp(i k^f_i a_i) \psi^f_\pm(x),
\nonumber\\
O:&\quad &^O\psi^\alpha_\pm(x) &= \mp \exp(\mp i \varphi(Ox)\pm i  
\tfrac{2 \pi}{3})\psi^\beta_\mp(Ox), \nonumber\\
&\quad &^O\psi^\beta_\pm(x)&= \mp \exp(\mp i \varphi(Ox)\mp i 
\tfrac{2 \pi}{3})\psi^\alpha_\mp(Ox), \nonumber \\
R:&\quad &^R\psi^\alpha_\pm(x) &= \psi^\beta_\pm(Rx), \quad\;\;\;
^R\psi^\beta_\pm(x) = \psi^\alpha_\pm(Rx), \nonumber \\
T:&\quad &^T\psi^\alpha_\pm(x) &= \exp(\mp i \varphi(Tx))
\psi^{\beta\dagger}_\pm(Tx),
\nonumber \\
&\quad &^T\psi^\beta_\pm(x) &= \exp(\mp i \varphi(Tx))
\psi^{\alpha\dagger}_\pm(Tx),
\nonumber \\
&\quad &^T\psi^{\alpha\dagger}_\pm(x) &= -\exp(\pm i \varphi(Tx)) 
\psi^\beta_\pm(Tx),\nonumber \\
&\quad &^T\psi^{\beta\dagger}_\pm(x) &= -\exp(\pm i \varphi(Tx)) 
\psi^\alpha_\pm(Tx).
\end{alignat}
Here $U(1)_Q$ is the fermion number symmetry of the holes. Interestingly, in 
the effective continuum theory the location of holes in lattice momentum space 
manifests itself as a ``charge'' $k^f_i$ under the displacement symmetry $D_i$.

Once the relevant low-energy degrees of freedom have been identified and the
transformation rules of the corresponding fields have been understood, the
construction of the effective action is uniquely determined. The low-energy 
effective action of magnons and holes is constructed as a derivative expansion.
At low energies, terms with a small number of derivatives dominate the 
dynamics. Since the holes are heavy nonrelativistic fermions, one 
time-derivative counts like two spatial derivatives. Here we limit ourselves to
terms with at most one temporal or two spatial derivatives. One then constructs
all terms consistent with the symmetries listed above. The effective action 
can be written as
\begin{equation}
S[\psi^{f\dagger}_\pm,\psi^f_\pm,\vec e] = \int d^2x \ dt \ \sum_{n_\psi}
{\cal L}_{n_\psi},
\end{equation}
where $n_\psi$ denotes the number of fermion fields that the various terms 
contain. The leading terms in the pure magnon sector take the form
\begin{eqnarray}
{\cal L}_0&=&\frac{\rho_s}{2} \left(\p_i \vec e \cdot \p_i \vec e  + 
\frac{1}{c^2} \p_t \vec e \cdot \p_t \vec e \right) \nonumber \\
&=&
2 \rho_s \left(v_i^+ v_i^- + \frac{1}{c^2} v_t^+ v_t^-\right).
\end{eqnarray}
The leading terms with two fermion fields (containing at most one temporal or 
two spatial derivatives) are given by
\begin{alignat}{2}
\label{Lagrangian2}
&&{\cal L}_2\,=\sum_{\ontopof{f=\alpha,\beta}{\, s = +,-}}& \Big[
M \psi^{f\dagger}_s \psi^f_s + \psi^{f\dagger}_s D_t \psi^f_s
+\frac{1}{2 M'} D_i \psi^{f\dagger}_s D_i \psi^f_s \nonumber \\
&\,&+\,&\Lambda \psi^{f\dagger}_s (i s v^s_1 + \sigma_f v^s_2) \psi^f_ 
{-s}\nonumber \\
&\,&+\,&iK \big[(D_1 + i s \sigma_f D_2) \psi^{f\dagger}_s
(v^s_1 + i s \sigma_f v^s_2)\psi^f_{-s} \nonumber \\
&\,&&-(v^s_1 + i s \sigma_f v^s_2)\psi^{f\dagger}_s (D_1 + i s  
\sigma_f D_2) \psi^f_{-s} \big] \nonumber \\
&\,&+\,&\sigma_f L \psi^{f\dagger}_s \epsilon_{ij}f^3_{ij}\psi^f_s +  
N_1 \psi^{f\dagger}_s v^s_i v^{-s}_i \psi^f_s \nonumber \\
&\,&+\,&i s \sigma_f N_2 \big( \psi^{f\dagger}_s v^s_1 v^{-s}_2  
\psi^f_s -
\psi^{f\dagger}_s v^s_2 v^{-s}_1 \psi^f_s \big) \Big].
\end{alignat}
Note that all low-energy parameters that appear above take real values. It 
should be noted that $v_i^\pm(x)$ contains one spatial derivative, such that
magnons and holes are indeed derivatively coupled. In eq.({\ref{Lagrangian2}), 
$M$ is the rest mass and $M'$ is the kinetic mass of a hole, $\Lambda$ is the
leading and $K$ is a subleading hole-one-magnon coupling, $L$, $N_1$ and $N_2$ 
are hole-two-magnon couplings, and
\begin{equation}
f^3_{ij}(x) = \p_i v^3_j(x) - \p_j v^3_i(x)
\end{equation}
is the field strength of the composite Abelian ``gauge'' field. The sign 
$\sigma_f$ is $+$ for $f = \alpha$ and $-$ for $f = \beta$. The covariant 
derivative in eq.({\ref{Lagrangian2}) takes the form
\begin{equation}
D_\mu \psi^f_\pm(x) = \p_\mu \psi^f_\pm(x) \pm i v^3_\mu(x) \psi^f_\pm(x).
\end{equation}
The leading terms with four fermion fields and without derivatives are given by
\begin{eqnarray}
\label{Lagrange4}
{\cal L}_4&=& \sum_{s = +,-} \Big\{
\frac{G_1}{2} (\psi^{\alpha\dagger}_s \psi^\alpha_s 
\psi^{\alpha\dagger}_{-s} \psi^\alpha_{-s} + 
\psi^{\beta\dagger}_s \psi^\beta_s 
\psi^{\beta\dagger}_{-s} \psi^\beta_{-s}) \nonumber \\
&+&G_2 \psi^{\alpha\dagger}_s \psi^\alpha_s \psi^{\beta\dagger}_s \psi^\beta_s 
+ G_3 \psi^{\alpha\dagger}_s \psi^\alpha_s 
\psi^{\beta\dagger}_{-s} \psi^\beta_{-s}\Big\}
\end{eqnarray}
with the real-valued 4-fermion coupling constants $G_1$, $G_2$, and $G_3$. 
In principle, there are even more contact interactions among the fermions, such
as 6- and 8-fermion couplings as well as 4-fermion couplings including 
derivatives. Since these terms play no role in the present work, we will not 
list them explicitly. 

Remarkably, the leading terms of the above Lagrangian have an accidental 
continuous $O(\gamma)$ rotation symmetry that acts as
\begin{eqnarray}
&&^{O(\gamma)}\psi^f_\pm(x) = \exp(\mp i \gamma/2)
\psi^f_\pm(O(\gamma)x),\nonumber \\
&&^{O(\gamma)} v_1(x) = \cos\gamma \ v_1(O(\gamma)x) + \sin\gamma \ 
v_2(O(\gamma)x), \nonumber \\
&&^{O(\gamma)} v_2(x) = - \sin\gamma \ v_1(O(\gamma)x) + \cos\gamma \ 
v_2(O(\gamma)x), \nonumber \\
&&O(\gamma)x = O(\gamma)(x_1,x_2,t) = \nonumber \\ 
&&(\cos\gamma \ x_1 - \sin\gamma \ x_2,\sin\gamma \ x_1 + \cos\gamma \ x_2,t).
\end{eqnarray}
This symmetry is not present in the underlying microscopic systems and is 
indeed explicitly broken by the higher-order terms in the effective action.

\section{Homogeneous versus Spiral Phases}
\label{2}
This section is devoted to the analysis of homogeneous and spiral 
configurations of the staggered magnetization, illustrated in figures 4 and 5,
respectively.
\begin{figure}[t]
\begin{center}
\epsfig{file=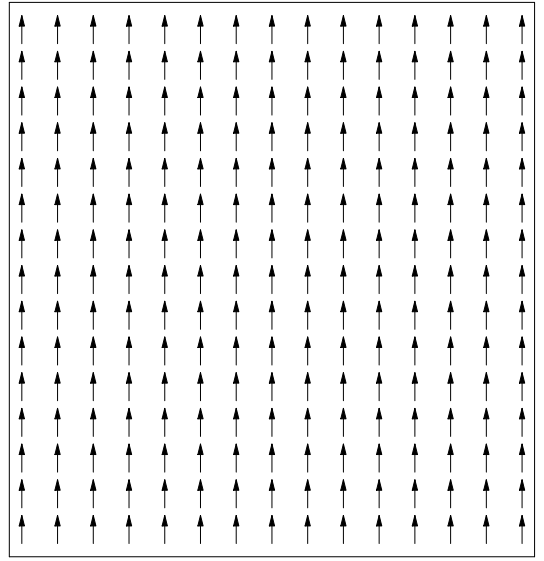,width=7cm}
\end{center}
\caption{\it Homogeneous phase with constant staggered magnetization.}
\end{figure}
\begin{figure}[t]
\begin{center}
\epsfig{file=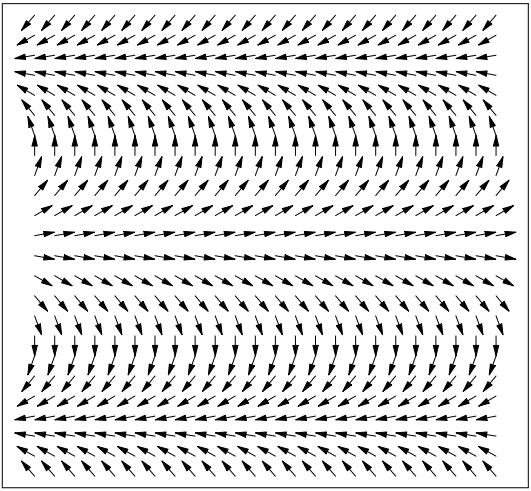,width=7cm}
\end{center}
\caption{\it Spiral phase with helical structure in the staggered
magnetization.}
\end{figure}
First, the energy of doped holes
is calculated keeping the staggered magnetization field fixed. Then the 
parameters of the staggered magnetization field are varied in order to minimize
the total energy.

\subsection{Fermionic Contribution to the Energy}

In this subsection we compute the fermionic contribution to the energy of a
homogeneous or spiral configuration of the staggered magnetization. For the 
moment, we ignore the 4-fermion couplings. The considerations of this paper are
valid only if the 4-fermion couplings are weak and can be treated in 
perturbation theory. Furthermore, we may neglect the vertices proportional to 
$K$, $L$, $N_1$, and $N_2$ which involve two spatial derivatives and
are thus of higher order than the hole-one-magnon vertex proportional to 
$\Lambda$. The fermion Hamiltonian resulting from the leading terms of the 
Euclidean action is given by
\begin{eqnarray}
H&=&\int d^2x \sum_{\ontopof{f=\alpha,\beta}{\, s = +,-}} 
\Big[ M \Psi^{f\dagger}_s \Psi^f_s +
\frac{1}{2 M'} D_i \Psi^{f\dagger}_s D_i \Psi^f_s \nonumber \\
&+&\Lambda \Psi^{f\dagger}_s(i s v^{s}_{1} + \sigma_f v^{s}_{2}) \Psi^f_{-s} 
\Big].
\end{eqnarray}
The covariant derivative takes the form
\begin{equation}
D_i \Psi^f_\pm(x) = \p_i \Psi^f_\pm(x) \pm i v^3_i(x) \Psi^f_\pm(x).
\end{equation}
Here $\Psi^{f\dagger}_s(x)$ and $\Psi^f_s(x)$ are creation and annihilation 
operators (not Grassmann numbers) for fermions of flavor $f = \alpha,\beta$ and
spin $s = +,-$ (parallel or antiparallel to the local staggered magnetization),
which obey canonical anticommutation relations. As before, $\sigma_\alpha = 1$ 
and $\sigma_\beta = - 1$. The above Hamiltonian is invariant against 
time-independent $U(1)_s$ gauge transformations
\begin{eqnarray}
\Psi^f_\pm(x)'&=&\exp(\pm i \alpha(x)) \Psi^f_\pm(x), \nonumber \\
{v^3_i}(x)'&=&v^3_i(x) - \p_i \alpha(x), \nonumber \\
{v^\pm_i}(x)'&=&v^\pm_i(x) \exp(\pm 2 i \alpha(x)).
\end{eqnarray}

Here we consider holes propagating in the background of a configuration with
\begin{equation}
{v^3_i}(x)' = c^3_i, \qquad {v^\pm_i}(x)' = c_i \in \R,
\end{equation}
where $c^{3}_i$ and $c_i$ are real-valued constants. In other words, we focus 
on configurations of the staggered magnetization in which (after an appropriate
gauge transformation) the fermions experience a constant composite vector field
${v_i}(x)'$, which leads to a homogeneous fermion density. As was shown in 
\cite{Bru07}, the most general configuration of this kind represents a spiral
in the staggered magnetization. The Hamiltonian can then be diagonalized by 
going to momentum space. Since magnon exchange does not mix the flavors, the 
Hamiltonian can be considered separately for $f = \alpha$ and $f = \beta$, but 
it still mixes spin $s = +$ with $s = -$. The single-particle Hamiltonian for 
holes with spatial momentum $\vec p = (p_1,p_2)$ takes the form
\begin{equation}
\label{Hf}
H^f(\vec p) = \left(\begin{array}{cc}
M + \frac{(p_i - c_i^3)^2}{2 M'} &
\Lambda (i c_1 + \sigma_f c_2) \\ \Lambda (- i c_1 + \sigma_f c_2) &
M + \frac{(p_i + c_i^3)^2}{2 M'}  \end{array} \right).
\end{equation}
The hole-one-magnon vertex proportional to $\Lambda$ mixes the spin $s = +$ and
$s = -$ states and provides a potential mechanism to stabilize a spiral phase.
The diagonalization of the above Hamiltonian yields
\begin{align}
\label{energy}
E^f_\pm(\vec p)=\,\, & M + \frac{p_i^2 + (c_i^3)^2}{2 M'}
\pm \sqrt{\left(\frac{p_i c_i^3}{M'}\right)^2 + \Lambda^2 |c|^2},
\end{align}
where $|c| = \sqrt{c^{2}_{1} + c^{2}_{2}}$. Interestingly, the above equation is 
independent of the flavor index $f$. We will keep the flavor index to indicate 
that there are two flavors in our calculations. Since the energy depends only
on $|c|$, unlike in the square lattice case, potential spiral configurations do
not prefer any particular spatial direction. This is due to the $O(\gamma)$ 
spatial rotation symmetry discussed in the previous section. However, one 
should keep in mind that $O(\gamma)$ is an accidental symmetry of just the 
leading terms in the effective action, which is broken explicitly by the 
higher-order terms. Hence, when the higher-order terms are included, one 
expects the spiral to align with a lattice direction. Mixing via the $\Lambda$ 
vertex lowers the energy $E^f_-$ and raises the energy $E^f_+$. It should be 
noted that, in this case, the index $\pm$ no longer refers to the spin 
orientation. Indeed, the eigenvectors corresponding to $E^f_\pm$ are linear 
combinations of both spins. The minimum of the energy is located at 
$\vec p = 0$ for which
\begin{equation}
E^f_\pm(0) = M + \frac{(c_i^3)^2}{2 M'} \pm \Lambda |c|.
\end{equation}
Since $c_i^3$ does not affect the magnon contribution to the energy density, we
fix it by minimizing $E^f_-(0)$ which implies $c_1^3 = c_2^3 = 0$. The energies
of eq.(\ref{energy}) then reduce to
\begin{equation}
E^f_\pm(\vec p) = M + \frac{p_i^2}{2 M'} \pm \Lambda |c|.
\end{equation}
Consequently, the filled hole pockets $P^f_\pm$ are circles determined by
\begin{equation}
\frac{p_i^2}{2 M'} = T^f_\pm,
\end{equation}
where $T^f_\pm$ is the kinetic energy of a hole in the pocket $P^f_\pm$ at the 
Fermi surface. The area of an occupied hole pocket determines the fermion 
density as
\begin{equation}
n^f_\pm = \frac{1}{(2 \pi)^2} \int_{P^f_\pm} d^2p = \frac{1}{2 \pi} M' T^f_\pm.
\end{equation}
The kinetic energy density of a filled pocket is given by
\begin{equation}
t^f_\pm = \frac{1}{(2 \pi)^2} \int_{P^f_\pm} d^2p \ \frac{p_i^2}{2 M'} = 
\frac{1}{4 \pi} M' {T^f_\pm}^2.
\end{equation}
The total density of fermions of all flavors is
\begin{eqnarray}
n &=& n^\alpha_+ + n^\alpha_- + n^\beta_+ + n^\beta_- \nonumber \\
&=& \frac{1}{2 \pi}
M'(T^\alpha_+ + T^\alpha_- + T^\beta_+ + T^\beta_-),
\end{eqnarray}
and the total energy density of the holes is
\begin{equation}
\epsilon_h = \epsilon^\alpha_+ + \epsilon^\alpha_- + 
\epsilon^\beta_+ + \epsilon^\beta_-,
\end{equation}
with
\begin{equation}
\epsilon^f_\pm = (M \pm \Lambda |c|) n^f_\pm + t^f_\pm.
\end{equation}
The filling of the various hole pockets is controlled by the parameters
$T^f_\pm$ which must be varied in order to minimize the energy while keeping 
the total density of holes fixed. We thus introduce
\begin{equation}
S = \epsilon_h - \mu n,
\end{equation}
where $\mu$ is a Lagrange multiplier that fixes the density, and we demand
\begin{equation}
\label{mini}
\frac{\p S}{\p T^f_\pm} = \frac{1}{2 \pi} M'(M \pm \Lambda |c| + 
T^f_\pm - \mu) = 0.
\end{equation}

\subsection{Four Populated Hole Pockets}

We will now populate the various hole pockets with fermions. First, we keep the
configuration of the staggered magnetization fixed and we vary the $T^f_\pm$ in
order to minimize the energy of the fermions. Then we also vary the parameters 
$c_i$ of the staggered magnetization field in order to minimize the total 
energy. One must distinguish various cases, depending on how many hole pockets 
are populated with fermions. In this subsection, we consider the case of 
populating all four hole pockets (i.e.\ with both flavors $f = \alpha, \beta$ 
and with both energy indices $\pm$). In this case, eq.(\ref{mini}) implies
\begin{equation}
\label{Teff}
\mu = M + \frac{\pi n}{2 M'}, \qquad 
T^f_\pm = \frac{\pi n}{2 M'} \mp \Lambda |c|.
\end{equation}
The total energy density then takes the form
\begin{eqnarray}
\epsilon&=&\epsilon_0 + \epsilon_m + \epsilon_h \nonumber \\
&=& \epsilon_0 + 2 \rho_s |c|^2 + 
\epsilon^\alpha_+ + \epsilon^\alpha_- + \epsilon^\beta_+ + \epsilon^\beta_-
\nonumber \\
&=&\epsilon_0 + 2 \rho_s |c|^2 + M n + \frac{\pi n^2}{4 M'} - 
\frac{1}{\pi} M' \Lambda^2 |c|^2.
\end{eqnarray}
Here $\epsilon_0$ is the energy density of the system at half-filling. For 
$2 \pi \rho_s > M' \Lambda^2$ the energy is minimized for $c_i = 0$ and the 
configuration is thus homogeneous. The total energy density in the 
four-pocket case is then given by
\begin{equation}
\label{etothom}
\epsilon_4 = \epsilon_0 + M n + \frac{\pi n^2}{4 M'}.
\end{equation}
For $2 \pi \rho_s < M' \Lambda^2$, on the other hand, the energy is not 
bounded from below. In this case, $|c|$ seems to grow without bound. However, 
according to eq.(\ref{Teff}) this would lead to $T^f_+ < 0$ which is physically
meaningless. What really happens is that two pockets get completely emptied and
we are naturally led to the two-pocket case. Before turning to that case, for
completeness we first discuss the three-pocket case.

\subsection{Three Populated Hole Pockets}

We now populate only three pockets with holes: the two pockets with the lower 
energies $E^\alpha_-$ and $E^\beta_-$ as well as the pocket with the higher 
energy $E^\alpha_+$. Of course, alternatively one could also fill the 
$\beta_+$-pocket. We now obtain
\begin{eqnarray}
n&=& n^\alpha_+ + n^\alpha_- + n^\beta_- = 
\frac{1}{2 \pi} M'(T^\alpha_+ + T^\alpha_- + T^\beta_-),\nonumber \\ 
\epsilon_h &=& \epsilon^\alpha_+ + \epsilon^\alpha_- + \epsilon^\beta_-,
\end{eqnarray}
such that eq.(\ref{mini}) yields
\begin{eqnarray}
\mu&=&M + \frac{2 \pi n}{3 M'} - \frac{\Lambda}{3} |c|,\nonumber \\
T^\alpha_+&=&\frac{2 \pi n}{3 M'} - \frac{4 \Lambda}{3} |c|, \nonumber \\
T^\alpha_-&=&T^\beta_- = \frac{2 \pi n}{3 M'} + \frac{2 \Lambda}{3} |c|.
\end{eqnarray}
The total energy density then takes the form
\begin{eqnarray}
\label{e3}
\epsilon&=&\epsilon_0 + \epsilon_m + \epsilon_h = \epsilon_0 + 2 \rho_s |c|^2 + 
\epsilon^\alpha_+ + \epsilon^\alpha_- + \epsilon^\beta_- \nonumber \\
&=&\epsilon_0 + 2 \rho_s |c|^2 + \left(M - \frac{\Lambda}{3} |c|\right) n + 
\frac{\pi n^2}{3 M'} \nonumber \\
&&-\,\, \frac{2}{3 \pi} M' \Lambda^2 |c|^2.
\end{eqnarray}
For $3 \pi \rho_s > M' \Lambda^2$ the energy density is bounded from below 
and its minimum is located at
\begin{equation}
|c| = \frac{\pi}{4} \frac{\Lambda n}{3 \pi \rho_s - M' \Lambda^2}.
\end{equation}
The resulting energy density in the three-pocket case takes the form
\begin{equation}
\label{etot3}
\epsilon_3 = \epsilon_0 + M n + 
\frac{\pi}{3 M'} \left(1 - \frac{1}{8} \frac{M' \Lambda^2}
{3 \pi \rho_s - M' \Lambda^2}\right) n^2.
\end{equation}
It is energetically less favorable than the homogeneous phase because 
$\epsilon_3 > \epsilon_4$ for $2 \pi \rho_s > M' \Lambda^2$. For 
$2 \pi \rho_s < M' \Lambda^2$ one obtains $T^\alpha_+ < 0$ which is unphysical.
In fact, the $\alpha_+$-pocket is then completely emptied and we are again led 
to investigating the two-pocket case.

\subsection{Two Populated Hole Pockets}

We now populate only two pockets with holes. These are necessarily the pockets
with the lower energies $E^\alpha_-$ and $E^\beta_-$. In this case we have
\begin{equation}
n = n^\alpha_- + n^\beta_- = \frac{1}{2 \pi} M'(T^\alpha_- + T^\beta_-), 
\,\,\,\, \epsilon_h = \epsilon^\alpha_- + \epsilon^\beta_-,
\end{equation}
and thus eq.(\ref{mini}) now implies
\begin{equation}
\mu = M + \frac{\pi n}{M'} - \Lambda |c|, \,\,\,
T^\alpha_- = T^\beta_- = \frac{\pi n}{M'}.
\end{equation}
The total energy density then takes the form
\begin{eqnarray}
\label{e0degree}
\epsilon&=&\epsilon_0 + \epsilon_m + \epsilon_h = \epsilon_0 +
2 \rho_s |c|^2 + \epsilon^\alpha_- + \epsilon^\beta_- \nonumber \\
&=&\epsilon_0 + 2 \rho_s |c|^2 + \left(M - \Lambda |c|\right) n + 
\frac{\pi n^2}{2 M'}.
\end{eqnarray}
The energy density is bounded from below and has its minimum at
\begin{equation}
|c| = \frac{\Lambda}{4 \rho_s} n. 
\end{equation}
The value at the minimum is given by
\begin{equation}
\label{etotzero}
\epsilon_2 = \epsilon_0 + M n + 
\left(\frac{\pi}{2 M'} - \frac{\Lambda^2}{8 \rho_s}\right) n^2.
\end{equation}
The two-pocket spiral phase is less stable than the homogeneous phase if 
$\epsilon_2 > \epsilon_4$, which is the case for $2 \pi \rho_s > M' \Lambda^2$.
As we have seen, both the three- and the four-pocket calculation become
meaningless for $2 \pi \rho_s < M' \Lambda^2$, because the kinetic energies
$T^f_+$ then become negative which is unphysical. The two-pocket calculation, 
on the other hand, continues to make sense for $2 \pi \rho_s < M' \Lambda^2$.

\subsection{One Populated Hole Pocket}

Finally, let us populate only one hole pocket, say the states with energy 
$E^\alpha_-$. Of course, alternatively one could also occupy the 
$\beta_-$-pocket. One now obtains
\begin{equation}
T^\alpha_- = \frac{2 \pi n}{M'}.
\end{equation}
The total energy density then takes the form
\begin{eqnarray}
\epsilon&=&\epsilon_0 + \epsilon_m + \epsilon_h = \epsilon_0 +
2 \rho_s |c|^2 + \epsilon^\alpha_- \nonumber \\
&=&\epsilon_0 + 2 \rho_s |c|^2 + (M - \Lambda |c|) n + \frac{\pi n^2}{M'},
\end{eqnarray}
which is minimized for
\begin{equation}
|c| = \frac{\Lambda}{4 \rho_s} n,
\end{equation}
and the corresponding energy density takes the form
\begin{equation}
\label{etot45}
\epsilon_1 = \epsilon_0 + M n + 
\left(\frac{\pi}{M'} - \frac{\Lambda^2}{8 \rho_s}\right) n^2.
\end{equation}
The one-pocket spiral is always energetically less favorable than the 
two-pocket spiral. 

\section{Inclusion of 4-Fermion Couplings in  Perturbation Theory}
\label{3}
In this section the 4-fermion contact interactions are incorporated in 
perturbation theory. Depending on the microscopic system in question, the
4-fermion couplings may or may not be small. If they are large, the result of
the perturbative calculation should not be trusted. In that case, one could
still perform a variational calculation. In this work we limit ourselves to
first order perturbation theory. We will distinguish four cases: the 
homogeneous phase, the three-pocket spiral, the two-pocket spiral, 
and the one-pocket spiral. Finally, depending on the values of the 
low-energy parameters, we determine which phase is energetically favorable.

\subsection{Four-Pocket Case}

Let us first consider the homogeneous phase. The perturbation of the 
Hamiltonian due to the leading 4-fermion contact terms is given by
\begin{eqnarray}
\label{DeltaH}
\Delta H &=& \int d^2x \sum_{s = +,-} \Big[
\frac{G_1}{2} (\Psi^{\alpha\dagger}_s \Psi^\alpha_s 
\Psi^{\alpha\dagger}_{-s} \Psi^\alpha_{-s} \nonumber \\
&+& 
\Psi^{\beta\dagger}_s \Psi^\beta_s 
\Psi^{\beta\dagger}_{-s} \Psi^\beta_{-s})+G_2 \Psi^{\alpha\dagger}_s
\Psi^\alpha_s \Psi^{\beta\dagger}_s \Psi^\beta_s \nonumber \\
&+&
G_3 \Psi^{\alpha\dagger}_s \Psi^\alpha_s 
\Psi^{\beta\dagger}_{-s} \Psi^\beta_{-s} \Big].
\end{eqnarray}
It should be noted that $\Psi^{f\dagger}_s(x)$ and $\Psi^f_s(x)$ again are 
fermion creation and annihilation operators (and not Grassmann numbers). In the
homogeneous phase the fermion density is equally distributed among the two spin
orientations and the two flavors such that
\begin{equation}
\langle \Psi^{\alpha\dagger}_+ \Psi^\alpha_+ \rangle =
\langle \Psi^{\alpha\dagger}_- \Psi^\alpha_- \rangle = 
\langle \Psi^{\beta\dagger}_+ \Psi^\beta_+ \rangle =
\langle \Psi^{\beta\dagger}_- \Psi^\beta_- \rangle = \frac{n}{4}.
\end{equation}
The brackets denote expectation values in the unperturbed state. Since the 
fermions are uncorrelated, for $f \neq f'$ or $s \neq s'$ one has
\begin{equation}
\langle \Psi^{f\dagger}_s \Psi^f_s \Psi^{f'\dagger}_{s'} \Psi^{f'}_{s'} \rangle
= \langle \Psi^{f\dagger}_s \Psi^f_s \rangle 
\langle \Psi^{f'\dagger}_{s'} \Psi^{f'}_{s'}\rangle.
\end{equation}
Taking the 4-fermion contact terms into account in first order perturbation
theory, the total energy density of eq.(\ref{etothom}) receives an additional 
contribution and now reads
\begin{equation}
\label{eps4}
\epsilon_4 = \epsilon_0 + M n + \frac{\pi n^2}{4 M'} +
\frac{1}{8} (G_1 + G_2 + G_3) n^2.
\end{equation}

\subsection{Three-Pocket Case}

For a spiral aligned along the $1$-direction ($c_1 > 0, c_2=0$) with 
$c^3_i = 0$ the eigenvectors of the single-particle Hamiltonian of 
eq.(\ref{Hf}) corresponding to the energy eigenvalues $E^f_\pm(\vec p)$ are 
given by
\begin{eqnarray}
\widetilde \Psi^f_\pm &=& \frac{1}{\sqrt{2}}(\Psi^f_- \pm i \Psi^f_+) \ 
\Rightarrow \nonumber \\
\Psi^f_{-}&=&\frac{1}{\sqrt{2}}(\widetilde \Psi^f_+ + \widetilde \Psi^f_-),
\,\,\,
\Psi^f_{+} = \frac{1}{\sqrt{2} i}(\widetilde \Psi^f_+ -
\widetilde \Psi^f_-).
\end{eqnarray}
Inserting this expression in eq.(\ref{DeltaH}) allows us to evaluate the
expectation value $\langle \Delta H \rangle$ in the unperturbed states
determined before. In the three-pocket case the states 
with energies $E^\alpha_-(\vec p)$, $E^\beta_-(\vec p)$, as well as 
$E^\alpha_+(\vec p)$ (or alternatively $E^\beta_+(\vec p)$), and with $\vec p$ 
inside the respective hole pocket are occupied and one arrives at
\begin{eqnarray}
&&\!\!\!\!\!\langle \widetilde \Psi^{\alpha\dagger}_+ \widetilde 
\Psi^\alpha_+ \rangle = \left(1 - \frac{1}{2} \frac{M' \Lambda^2}
{3 \pi \rho_s - M' \Lambda^2}\right) \frac{n}{3}, \,\,\,\,
\langle \widetilde \Psi^{\beta\dagger}_+ \widetilde \Psi^\beta_+ \rangle = 0,
\nonumber \\
&&\!\!\!\!\!\langle \widetilde \Psi^{\alpha\dagger}_- \widetilde 
\Psi^\alpha_- \rangle =
\langle \widetilde \Psi^{\beta\dagger}_- \widetilde \Psi^\beta_- \rangle =
\left(1 + \frac{1}{4} \frac{M' \Lambda^2}{3 \pi \rho_s - M' \Lambda^2}\right) 
\frac{n}{3}.
\end{eqnarray}
As a result, the energy density of eq.(\ref{etot3}) turns into
\begin{eqnarray}
\label{eps3}
\epsilon_3&=&\epsilon_0 + M n + 
\frac{\pi}{3 M'}\left(1 - \frac{1}{8}
\frac{M' \Lambda^2}{3 \pi \rho_s - M' \Lambda^2}\right) n^2 \nonumber \\
&+&\frac{4 \pi \rho_s - M' \Lambda^2}{(3 \pi \rho_s - M' \Lambda^2)^2} 
\frac{1}{32} \Big[8 (G_1 + G_2 + G_3) \pi \rho_s \Big. \nonumber \\
&-&\Big. (4 G_1 + 3 G_2 + 3 G_3) M' \Lambda^2\Big] n^2.
\end{eqnarray}

\subsection{Two-Pocket Case}

In this case only the states with energy $E^\alpha_-(\vec p)$ and 
$E^\beta_-(\vec p)$ with $\vec p$ inside the respective hole pocket $P^f_-$ are
occupied and hence
\begin{equation}
\langle \widetilde \Psi^{\alpha\dagger}_- \widetilde \Psi^\alpha_- \rangle =
\langle \widetilde \Psi^{\beta\dagger}_- \widetilde \Psi^\beta_- \rangle =
\frac{n}{2}, \,\,\,\,
\langle \widetilde \Psi^{\alpha\dagger}_+ \widetilde \Psi^\alpha_+ \rangle =
\langle \widetilde \Psi^{\beta\dagger}_+ \widetilde \Psi^\beta_+ \rangle = 0.
\end{equation}
As a result the energy density of eq.(\ref{etotzero}) turns into
\begin{equation}
\label{eps2}
\epsilon_2 = \epsilon_0 + M n + \left(\frac{\pi}{2 M'} - 
\frac{\Lambda^2}{8 \rho_s}\right) n^2 + \frac{1}{8} (G_2 + G_3) n^2.
\end{equation}

\subsection{One-Pocket Case}

In the one-pocket case only the states with energy $E^\alpha_-(\vec p)$ 
(or alternatively with $E^\beta_-(\vec p)$) and with $\vec p$ inside the
corresponding hole pocket are occupied so that one has
\begin{equation}
\langle \widetilde \Psi^{\alpha\dagger}_- \widetilde \Psi^\alpha_- \rangle = n,
\,\,\,\,
\langle \widetilde \Psi^{\alpha\dagger}_+ \widetilde \Psi^\alpha_+ \rangle =
\langle \widetilde \Psi^{\beta\dagger}_+ \widetilde \Psi^\beta_+ \rangle = 
\langle \widetilde \Psi^{\beta\dagger}_- \widetilde \Psi^\beta_- \rangle = 0.
\end{equation}
In this case, the 4-fermion terms do not contribute to the energy density which
thus maintains the form of eq.(\ref{etot45}), i.e.
\begin{equation}
\label{eps1}
\epsilon_1 = \epsilon_0 + M n + 
\left(\frac{\pi}{M'} - \frac{\Lambda^2}{8 \rho_s}\right) n^2.
\end{equation}

\subsection{Stability Ranges of Various Phases}

Let us summarize the results of the previous subsections. The energy densities 
of the various phases take the form
\begin{equation}
\epsilon_i = \epsilon_0 + M n + \frac{1}{2} \kappa_i n^2.
\end{equation}
According to eqs.(\ref{eps1}), (\ref{eps2}), (\ref{eps3}), and (\ref{eps4}), 
the compressibilities $\kappa_i$ are given by
\begin{eqnarray}
\kappa_1&=&\frac{2 \pi}{M'} - \frac{\Lambda^2}{4 \rho_s}, \nonumber \\
\kappa_2&=&\frac{\pi}{M'} - \frac{\Lambda^2}{4 \rho_s}
+ \frac{1}{4} (G_2 + G_3), \nonumber \\
\kappa_3&=&\frac{2 \pi}{3 M'} \left(1 - \frac{1}{8}\frac{M' \Lambda^2}
{3 \pi \rho_s - M' \Lambda^2}\right) \nonumber \\
&+&\frac{4 \pi \rho_s - M' \Lambda^2}{(3 \pi \rho_s - M' \Lambda^2)^2}
\frac{1}{16} \Big[8(G_1 + G_2 + G_3)\pi\rho_s \Big. \nonumber \\
&-&\Big. (4 G_1+ 3 G_2 + 3 G_3) M' \Lambda^2\Big], \nonumber \\
\kappa_4&=&\frac{\pi}{2 M'} + \frac{1}{4} (G_1 + G_2 + G_3).
\end{eqnarray}
The compressibilities $\kappa_i$ as functions of $M'\Lambda^2/2\pi\rho_s$ are
shown in figure 6. 
\begin{figure}[t]
\begin{center}
\epsfig{file=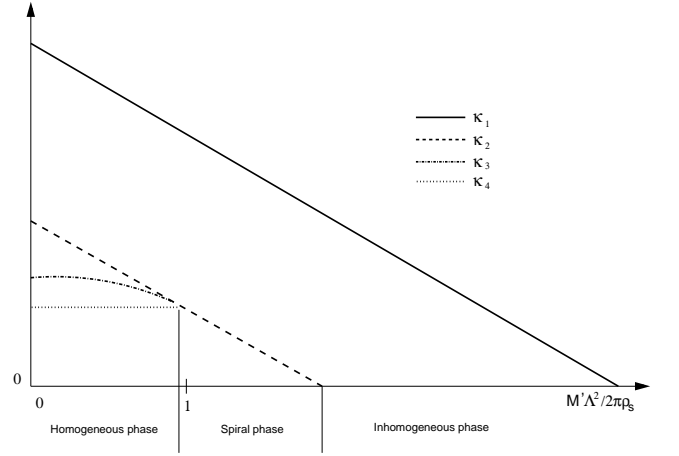,width=8.5cm}
\end{center}
\caption{\it The compressibilities $\kappa_i$ as functions of 
$M' \Lambda^2/2 \pi \rho_s$ determine the stability ranges of the various 
phases. A homogeneous phase, a spiral, or an inhomogeneous phase 
are energetically favorable, for large, intermediate, and small values of 
$\rho_s$, respectively.}
\end{figure}
For large values of $\rho_s$, spiral phases cost a large amount of magnetic 
energy and the homogeneous phase is more stable. To be more precise, in this 
regime one has 
$\kappa_4 < \kappa_3 < \kappa_2 < \kappa_1$. Notice that 
$\kappa_1$ is always larger than $\kappa_2$ for any value of $\rho_s$. As 
$\rho_s$ decreases and reaches the value
\begin{equation}
\rho_s = \frac{M'\Lambda^2}{2\pi} + \frac{(M')^{2}\Lambda^2G_1}{4\pi^2},
\end{equation}
at leading order in the 4-fermi couplings one finds 
$\kappa_2 = \kappa_3 = \kappa_4$. For smaller values of $\rho_s$, the 
two-pocket spiral is energetically favored until $\kappa_2$ becomes negative 
and the system becomes unstable against the formation of spatial 
inhomogeneities of a yet undetermined type.

It should be pointed out again that these results apply only if the 4-fermion 
contact interactions are weak. Even if the 4-fermion couplings are indeed
small, the results presented in this work do not necessarily reveal the true 
nature of the ground state. Due to the variational nature of the calculation, 
one cannot exclude that the phases that we found may still be unstable in 
certain parameter regions. It is instructive to compare the results presented 
here with the results obtained in the square lattice case \cite{Bru07}. 
Qualitatively the stability ranges of various phases are the same for both 
lattice geometries except that the one-pocket spiral is never energetically 
favored on the honeycomb lattice while it is favorable in a small parameter 
regime on the square lattice.  

\section{Conclusions and Outlook}
\label{4}
In this paper we have used a systematic effective field theory for 
antiferromagnetic magnons and holes on the honeycomb lattice to investigate the
dynamics of holes in the background of a staggered magnetization field. We
have limited ourselves to constant composite vector fields $v_i(x)'$ which
implies that the fermions experience a constant background field. 
Interestingly, unlike in the square lattice case, due to the accidental 
continuous $O(\gamma)$ spatial rotation symmetry, at leading order a spiral 
does not have an a priori preferred spatial direction. However, 
since the $O(\gamma)$ symmetry is broken explicitly by the higher-order 
terms, once such terms are included, one expects the spiral to align with a 
lattice direction. Finally, we investigated the stability of spiral phases in 
the presence of 4-fermion couplings. Assuming that the 4-fermion couplings 
can be treated perturbatively, we have seen that, for sufficiently large 
values of $\rho_s$, the homogeneous phase is energetically favored. With 
decreasing $\rho_s$, a two-pocket spiral becomes energetically more favorable. 
On the other hand, in contrast to the square lattice case, the one-pocket 
spiral is never favored. For small values of $\rho_s$ the two-pocket spiral
becomes unstable against the formation of inhomogeneities of a yet undetermined
type. In \cite{Jia08} the low-energy parameters $\rho_s$, $c$, and $M'$ have 
been determined in terms of the parameters $t$ and $J$ of the underlying 
$t$-$J$ model. It will be interesting to also determine the strength of the 
hole-one-magnon vertex $\Lambda$ in order to decide which phase is realized in 
this model. Further applications of the effective theory, including the 
one-magnon exchange potential and the resulting two-hole bound states, are 
currently under investigation. \newline \newline

\section*{Acknowledgments}

We like to thank B.\ Bessire, M.\ Nyfeler, and M.\ Wirz for their collaboration
on the construction of the effective action for magnons and holes on the 
honeycomb lattice. U.-J.\ W. likes to thank P.\ A.\ Lee for interesting 
discussions and the members of the Center for Theoretical Physics at MIT, where
part of this work was performed, for their hospitality. C.\ P.\ H.\ would like 
to thank the members of the Institute for Theoretical Physics at Bern 
University for their hospitality. The work of C.\ P.\ H.\ is supported by
CONACYT Grant No.\ 50744-F and by Grant Proyecto Cuerpo-Academico-56-UCOL. This
work was also supported in part by funds provided by the Schweizerischer 
Nationalfonds (SNF). In particular, F.\ K.\ is supported by an SNF young 
researcher fellowship. The ``Center for Research and Education in Fundamental
Physics'' at Bern University is supported by the ``Innovations- und 
Kooperationsprojekt C-13'' of the Schweizerische Universit\"atskonferenz 
(SUK/CRUS).

\end{document}